\documentclass[aps,showpacs,prl]{revtex4}
\newcommand{\dslash}{D\!\!\!\!\slash}
\begin{document}

\title{Functional integrals for QCD at nonzero chemical potential and zero density}

\author{Thomas D. Cohen}
\email{cohen@physics.umd.edu}

\affiliation{Department of Physics, University of Maryland, College
Park, MD 20742-4111}

\begin{abstract}
In a Euclidean space functional integral treatment of the free
energy of QCD, a chemical potential enters only through the
functional determinant of the Dirac operator which for any flavor
is $\dslash + m - \mu_f \gamma_0$ (where $\mu_f$ is the chemical
potential for the given flavor).  Any nonzero $\mu$ alters all of
the eigenvalues of the Dirac operator relative to the $\mu=0$
value, leading to a naive expectation that the determinant is
altered and which thereby alters the free energy.
Phenomenologically, this does not occur at $T=0$ for sufficiently
small $\mu$, in contradiction to this naive expectation.   The
problem of how to understand this phenomenological behavior in
terms  functional integrals is solved for the case of an isospin
chemical through the study of the spectrum of the operator
$\gamma_0 (\dslash + m)$.  The case of the baryon chemical
potential is briefly discussed.
\end{abstract}


\maketitle

Consider QCD with two degenerate light flavors at zero temperature and a finite isospin chemical potential.   At a phenomenological level much is known about this system \cite{ss}.  The behavior at low chemical potential is of interest here. The energy density and isospin density are known to be zero for $|\mu_I| < m_\pi$.  As $|\mu_I|$ is increased above $m_\pi$ a second order phase transition to a pion condensed state is reached in which the energy density and isospin density both continuously increase from zero.  It is pretty clear how to interpret these results in terms of eigenstates of the QCD hamiltonian.  The critical value of $\mu_I$ corresponds to the state in the system with the  smallest energy per unit isospin (namely a pion at rest). Unfortunately, there is no simple way to obtain these eigenstates starting directly from the QCD.   Here the focus will be on understanding the phenomenological behavior in terms of a Euclidean space functional integral formulation of the theory.

For simplicity of presentation, the discussion will be restricted to two-flavored QCD with degenerate flavors.  The inclusion of additional heavy flavors is straightforward and does not alter any of the results.
As will be clear, it is necessary to work at finite temperature and then study the limit as $T \rightarrow \infty$ at a later stage. The free energy is given by $G_I(T, \mu_I)= E - T S -  \mu_I I_3$  where $E$, $T$, $S$ and $\mu_I$  are the energy, temperature,  entropy density, isospin and chemical potential respectively.  One can envision working in a finite but large box of volume $V$ with appropriate boundary conditions.   The thermodynamic limit $ V \rightarrow \infty$ can be taken at the end of the calculation and results may then be expressed  in terms of intensive variables such ${\cal E} = E/V$, ${\cal G} =G/V$  or $\rho_I =I_3/V$. The grand partition function, $Z_I(T \mu_I)= e^{-\beta G(T, \mu_I)}$, with $\beta =1/T$ can be represented as a functional integral,
\begin{equation}
Z_I(T,\mu_I) \, =
\int  d [A] \,  \, \left | {\rm det} \left ( \dslash + m - \frac{\mu_I \gamma_0}{2} \, \right ) \right |^2 e^{-S_{YM}} \label{funcint}
\end{equation}
where  $S_{YM}$ is the Yang-Mills action and the functional integral is evaluated over a four dimensional box with a temporal length of $\beta$ and with periodic boundary conditions for the gluons.  The functional determinant is computed over the same box but with anti-periodic boundary conditions for the fermions and is taken to be over a single flavor. The absolute value squared arises since the functional determinant of the up quarks is the complex conjugate of the functional determinant of the down quarks provided the chemical potentials for the two flavors are equal and opposite \cite{AlfKapWil99,Coh03}. Thus the only way that the presence of the chemical potential is felt is via the functional determinant of the Dirac operator.     The determinant may be represented as the product of the eigenvalues of the Dirac operator $
{\rm det}\left( \dslash + m - \frac{\mu_I  \gamma_0}{2} \,
 \,  \right )   =   \prod_j \lambda_j \; \; \;
 {\rm where} \; \; \; \left( \dslash + m - \frac{\mu_I \gamma_0}{2} \,
 \,  \right ) \psi_j  =  \lambda_j \psi_j$ .

At this stage one is confronted by something of a puzzle.
{\it A priori}, one expects that the eigenspectrum of the Dirac operator with $\mu_I =0$ (and some fixed gauge configuration) is completely different from the spectrum with any nonzero $\mu_I$ in the sense that every eigenvalue is expected to be different; there is no mathematical reason for the eigenvalues not to depend on $\mu$.  In the absence of some conspiracy between the eigenvalues, one would therefore expect that  the functional determinant with any nonzero $\mu_I$ would differ from  the functional determinant $\mu_I=0$ and that this would occur for every gauge configuration.   In turn,  eq.~(\ref{funcint}), would lead one to expect that if the functional determinants all differ from their  $\mu_I=0$ counterparts, that $Z_I(T,\mu_I)$ also would vary.  Moreover, this expectation would seem to hold at any temperature including $T=0$.  Clearly, this expectation is wrong.  At $T=0$ we know phenomenologically that the free energy is precisely its vacuum value for nonzero values of  $\mu_I$ provided that $|\mu_I| < m_\pi$.  The puzzle is simply how can we understand the vanishing physical effects of $|\mu_I| < m_\pi$ in the context of a functional integral treatment.

In one sense, this problem might be dismissed as, quite literally, much ado about nothing; the problem  amounts to getting a mathematical understanding of nothing happening at the physical level.  However, to understand the phase transition where something {\it does} occur, it is necessary to understand just how it is that nothing occurs below the transition. Accordingly, it seems fitting to name this the isospin ``Silver Blaze'' problem after the Arthur Conan Doyle story of that name in which the ``curious incident'' of a dog doing nothing in the night time provides Holmes with an essential clue. There are analogous Silver Blaze problems associated with the baryon chemical potential, the strangeness chemical potential or various linear combinations of these chemical potentials.

The key to the solution of the isospin Silver Blaze problem is the
study of eigenstates of $\gamma_0$ times the usual Dirac
operator.  The eigenspectrum of the usual Dirac operator,
$\dslash +m $, has received considerable attention in both random
matrix models\cite{rm} and lattice studies\cite{lat}; the intense
interest is because this eigenspectrum provides essential
insight  into chiral symmetry breaking\cite{bc}.  However
$\gamma_0$ times the Dirac operator has not received such
interest.  There have been some studies of the so-called
propagator matrix, which is a lattice version of  $\gamma_0$
times the Dirac operator in connection with the Glasgow
method\cite{Glasgow} and in a random matrix form\cite{rm2}.  Such
treatments are not completely illuminating for the present
purposes since the key result is manifest only in the continuum
limit (or strictly speaking where the continuum limit has been
taken at least in the time direction).  In this limit, the
eigenspectrum  of $\gamma_0$ times the Dirac operator provides
insight into the physics of pion condensation in much the same
way as the eigenspectrum of the usual Dirac operator gives
insight into chiral symmetry breaking.  To see how this comes
about consider the partition function of QCD in the presence of
an isospin chemical potential as given in eq.~(\ref{funcint})
with $\mu_I$ solely affecting the functional determinant.  It is
a simple exercise in linear algebra to show that functional
determinant can be represented as
\begin{equation}
{\rm det}\left ( \dslash+ m - \frac{\mu_I  \gamma_0}{2}
 \, \right )  =
 {\rm det}\left ( \dslash + m  \, \right ) \, e^{- \frac{1}{2} \int_0^{\mu_I} \, {\rm d} \mu_I' {\rm tr} \left[ \left( \gamma_0 (\dslash + m) -  \frac{\mu_I'}{2} \right )^{-1} \right]}
 \label{detform}
\end{equation}
which shows that all of the effects of the chemical potential can be represented as the exponential of an integral of the trace of $\gamma_0$ times the usual Dirac operator.  This trace can be evaluated via a sum over eigenvalues with the functions subject to anti-periodic boundary conditions.

 There are several features of the operator
$ \left( \gamma_0 (\dslash + m) -  \frac{\mu_I'}{2} \right) $
which greatly facilitate the computation of the trace. One useful
property concerns the paring of eigenvalues.  It is easily
demonstrated that if $\lambda$ is an eigenvalue of  $\left (
\gamma_0 (\dslash + m) \right)$, then so is $-\lambda^*$. Next
notice that the time derivative term in the operator is not
multiplying any $\gamma$ matrices since $\gamma_0^2 = 1$.
Moreover, $\mu_I'$ also does not multiply a $\gamma$ matrix but
simply enters as an additive constant and consequently has no
role in determining the eigenfunctions; it merely shifts the
eigenvalues.    Together these imply that if $   \left ( \gamma_0
(\dslash + m)  \right) |\psi_j \rangle = \lambda |\psi \rangle$,
then
 $ \left ( \gamma_0 (\dslash + m) - \frac{\mu_I'}{2} \right)
e^{i \omega t} |\psi \rangle = \left (\lambda - \frac{\mu_I'}{2}+
i \omega \right ) e^{i \omega t}|\psi \rangle$.  This means that
for any given eigenfunction, another can be generated with an
eigenvalue shifted by $i \omega$ by simply multiplying by a
time-dependent phase. Note at this stage it is critical that the
continuum expression is used for the time derivatives.
Moreover,  these phase factor cannot be arbitrary.  Recall that
the eigenfunctions are constrained to be anti-periodic in time
with a periodicity of $\beta$.  Thus, the eigenfunctions  of $
\left( \gamma_0 (\dslash + m) -  \frac{\mu_I'}{2} \right) $ can
be grouped into families denoted  by two indices an index $j$
representing an ``intrinsic'' eigenstate and an index $n$
representing the phase factor satisfying:
\begin{equation}
 |\psi_{j n+1} \rangle  =  e^{\frac{i 2 \pi  t}{\beta}}|\psi_{j n}  \rangle  \; \; \; \; \;
\lambda_{j n}  =  \epsilon_j - \frac{\mu_I'}{2} + i \left( \frac{\phi_j}{\beta} + \frac{(2 n +1 ) \pi}{\beta} \right ) \label{eps}
\end{equation}
where $\lambda_{j n}$, the eigenvalue of the operator, is broken up into a real and an imaginary part.  This decomposition is unique if we impose the condition that $-\pi \le \phi_j < \pi$.   The form of eq.~(\ref{eps}) is highly suggestive.  Indeed, if one ignores the fact that $\phi_j$ is generically nonzero, it is identical to the form for the case of free noninteracting fermions; in the noninteracting case, $\epsilon_j$ simply represents the energy of a mode.  Here, $\epsilon_j$, the real part of eigenstates of $\gamma_0$ times the usual Euclidean Dirac operator, may be considered as a quasi-energy.  Of course, the quasi-energies differ from the energies in the free fermion case in critical ways; they depend on the gauge field configuration and are not known analytically.

The trace can be evaluated by first summing over the $n$ for the
eigenvalues (which is analogous to a usual Matsubura
sum\cite{Matsubura} but here done for fields in the presence of a
time-dependent background); a sum over the intrinsic  $j$ with
$\epsilon_j \ge 0$ is done subsuently:
\begin{equation}
{\rm tr} \left [ \left( \gamma_0 (\dslash + m) -  \frac{\mu_I'}{2} \right)^{-1} \right]\, =
\sum_j  \theta (\epsilon_j) \frac{\beta}{2} \,\left[
      \tanh \left(\frac{\beta}{2}
           \left( \epsilon_j  - \frac{\mu_I'}{2}  \right)  +
          i \,\frac{\phi_j}{2} \right ) -\tanh \left(\frac{\beta}{2} \,
            \left( \epsilon_j  + \frac{\mu_I'}{2}  \right)  -
           i \,\frac{\phi_j}{2} \right )  \right]\label{trsf}
\end{equation}

Recall that the problem of interest is at zero temperature which corresponds to $\beta \rightarrow \infty$. As this limit is approached, the asymptotic behavior of the hyperbolic tangent can be used to simplify the result of eq.~(\ref{trsf}):
\begin{equation}
{\rm tr} \left [\left( \gamma_0 (\dslash + m) -  \frac{\mu_I'}{2} \right)^{-1} \right]\, =
 \sum_j  \theta (\epsilon_j) \, {\rm sign}(\mu_I') \, \left [ - \beta \,\theta (|\mu_I'| - 2 \epsilon_j  ) + i 2 \phi_j \, \delta (|\mu_I'| -  2 \epsilon_j ) \right ] + { \cal O}(e^{- \beta \Lambda}) \label{simpform}
\end{equation}
where $\Lambda$ represents a characteristic hadronic scale.  Note that
 this becomes exponentially small as $\beta \rightarrow \infty$ and will
 have a negligible effect.  Inserting eq.~(\ref{simpform}) into eq.~(\ref{detform}) yields
\begin{equation}
\frac{{\rm det}\left ( \dslash+ m - \frac{\mu_I  \gamma_0}{2}
 \, \right ) }{ {\rm det}\left ( \dslash + m  \, \right )} =
    \exp \left( - i \sum_j \phi_j \, \theta (\epsilon_j)  \,
     \theta (|\mu_I| -    2 \epsilon_j) \right )
 \exp \left( \frac{\beta }{2} \sum_j  \theta (\epsilon_j)  \,
 \theta (|\mu_I| - 2 \epsilon_j ) \, \left(|\mu_I| -2 \epsilon_j \right )
 + { \cal O}(e^{- \beta \Lambda}) \right )
 \label{finaldetform}
\end{equation}
In using eq.~(\ref{finaldetform}) rather than the full expression
in eq.~(\ref{trsf}) and neglecting the exponential, one is
implicitly assuming that the zero temperature limit for the
physical quantity of interest, the free energy, has a smooth
behavior as the zero temperature limit is approached.  This is
highly plausible at finite chemical poitential but has not been
directly proved from QCD.

Equation (\ref{finaldetform}) provides an essential ingredient for the  resolution of the isospin Silver Blaze problem.   The theta functions ensure that for any given gauge configuration in the QCD functional integral, as $T=0$ is approached, the functional determinant is unchanged from its $\mu_I=0$ value, unless $\frac{|\mu_I|}{2}$ is greater than the quasi-energy of the minimum positive quasi-energy mode.  Thus, eq.~(\ref{finaldetform}) explains how at zero temperature, functional determinants can remain unchanged despite all of the eigenvalues of the Dirac operator changing which is the core of the Silver Blaze problem.

Of course, the value of the minimum positive quasi-energy depends on the gauge configuration.  A full resolution of the isospin Silver Blaze problem depends on showing that nothing physical occurs at zero temperature unless
$\mu_I \ge m_\pi$.  Clearly this is the case, provided that the minimum positive quasi-energy is greater than or equal to $m_\pi$ for those configurations which contribute to the functional integral at zero temperature.  A formal statement of this condition can be given in terms of the spectral density (as a function of $\epsilon$ ) averaged over the gauge conditions.  We denote this spectral density as ${\hat\rho}(\epsilon)$
\begin{equation}
{\hat\rho}(\epsilon) = \sum_j \delta(\epsilon-\epsilon_j) = \frac{1}{2 \beta} \frac{\partial}{\partial \epsilon} {\rm tr} \left [   \left( \gamma_0 (\dslash + m) -  \epsilon \right)^{-1} + \left( (- \dslash + m)\gamma_0  -  \epsilon \right)^{-1} \right] + { \cal O}(e^{- \beta \Lambda})
\label{specden} \end{equation}
where $\epsilon_j$  is the $j^{\rm th}$ quasi-energy  for a given  configuration, and the second equality follows from eq.~(\ref{simpform}).   We denote averaging over gauge configurations with the notation:
\begin{equation}
\langle \hat{{\cal O}} \rangle_{T, \mu_I} =  \frac{1}{ Z_I(T,\mu_I)} \,
\int  d [A] \, \hat{{\cal O}} \, \left | {\rm det} \left ( \dslash + m - \frac{\mu_I \gamma_0}{2} \, \right ) \right |^2 e^{-S_{YM}}
\label{av} \end{equation}
where the functional integral is evaluated with standard boundary conditions.  The minimum relevant positive quasi-energy, $\epsilon_{\rm min}$, can be defined in the following way: $\langle \hat{\rho}(\epsilon) \rangle_{0 , 0} = 0 $ if and only if $|\epsilon | < \epsilon_{\rm min} $.
The Silver Blaze problem is then fully resolved if two conditions are met:
\begin{eqnarray}
i) &{}& \; \; \langle \hat{\rho}(\epsilon) \rangle_{0, \mu_I} =  \langle \hat{\rho}(\epsilon)\rangle_{0, 0} \; \; {\rm for \; all } \; \; \mu_I < 2 \epsilon_{\rm min} \nonumber \\
ii) &{}&\epsilon_{\rm min} = \frac{m_\pi}{2} \; .  \nonumber
 \end{eqnarray}
To verify that these conditions are sufficient, use the definition of the free energy in terms of $Z_I$ and eqs.~(\ref{funcint})) and (\ref{specden})  to note that $
 \frac{\partial G(0,\mu_I) }{\partial \mu} =  2 \int_0^{\frac{\mu_I}{2}} {\rm d} \epsilon  \, \langle \hat{\rho} (\epsilon)\rangle_{ 0,\mu_I}$.  Thus if $T=0$, $|\mu_I| < m_\pi$ and the two conditions are satisfied,  then free energy is independent of $\mu_I$ and the expectation value of the isospin vanishes .

It is straightforward to demonstrate that the two conditions are in fact met in QCD, provided that one assumes that no first order phase transition occurs for $T=0$ and $|\mu_I| < m_\pi$.  This assumption is innocuous---it is highly plausible  {\it a priori} and is known not to occur in nature.
To demonstrate  the validity of i) use eqs. (\ref{specden}) and (\ref{av}) to show that
\begin{equation}
     \frac{\partial}{\partial \mu_I} \langle \hat{\rho}(\epsilon) \rangle_{0, \mu_I} = \left \langle \hat{\rho}(\epsilon) \left ( \int_0^{\frac{\mu_I}{2}} {\rm d} \epsilon'  \hat{\rho}(\epsilon') \right ) \right \rangle_{0 , \mu_I}
\label{div} \end{equation}
Note that, the operator $\int_0^{\frac{\mu_I}{2}} {\rm d} \epsilon'  \hat{\rho}(\epsilon')$ is manifestly nonnegative as is the measure in eq.~(\ref{av}).  In general, if an operator $\hat{A}$ is manifestly nonnegative and it is averaged over a nonnegative measure and if $\langle
 \hat{A} \rangle = 0 $ it follows that $\langle   \hat{B} \hat{A} \rangle = 0 $.  Equation (\ref{div}) and this general condition  together imply that
$\frac{\partial}{\partial \mu_I} \langle \hat{\rho}(\epsilon) \rangle_{0, \mu_I} = 0 $ if  $\left \langle  \int_0^{\frac{\mu_I}{2}} {\rm d} \epsilon'  \hat{\rho}(\epsilon') \right \rangle_{0 , \mu_I}$ =0. This in turn implies  the validity of condition i) provided that $ \langle \hat{\rho}(\epsilon) \rangle_{0, \mu_I}$ does not change discontinuously  at some value of $\mu_I$ ({\it i.e.}, there is  no first order transition).

 To establish conditon ii,  consider the charged pseudo-scalar susceptibility $\chi_{\rm ps}^+ = \int {\rm d}^4 x \langle J_{-}(x) J_{+}(0)  \rangle$ (with $J_{+}=\overline{d} \gamma_5 u$) as a function of $\mu_I$.   This susceptibility can be represented as a functional integral; a straightforward computation using techniques similar to those used above, enables one to express this functional integral in terms of the spectrum of $\gamma_0 (\dslash + m)$:
\begin{equation}
 \chi_{\rm ps}^+(T, \mu_I)   = \frac{1}{ V }  \int {\rm d}\epsilon \, {\rm d}\phi \,  \frac{\left \langle \left (\sum_j \delta(\epsilon-\epsilon_j) \delta(\phi-\phi_j)  \right ) \right \rangle_{T, \mu_I} \, \sinh (\beta \,
      \left( \epsilon  - \frac{\mu_I }{2} \right) )}{
    \left( 2\,\epsilon  - \mu_I  \right) \,
    \left( \cosh (\beta \,
        \left( \epsilon  - \frac{\mu_I }{2} \right) ) +
      \cosh (\phi ) \right) }
      = \frac{1}{ V}\int {\rm d}\epsilon \, \,  \frac{\langle \hat{ \rho}(\epsilon)_{T , \mu_I} \left ( 1+ {\cal O}(e^{-\beta \Lambda}) \right)}{
    \left | 2\,\epsilon  - \mu_I  \right| }
    \label{chif}
\end{equation}
where the first equality is exact and the second equality exploits the fact that our interest is in the limit as $T \rightarrow 0$.  From the second equality in eq.~(\ref{chif})  we see that in the $T \rightarrow 0$ limit, $\chi_{\rm ps}^+$ diverges when $\frac{\mu_I}{2}$ is increased to the smallest value of $\epsilon$ for which the $\langle\hat{\rho}(\epsilon)\rangle_{ 0, \mu_I}$ is nonzero.    From condition i) (previously demonstrated to hold) this implies that
$\chi_{\rm ps}^+$ diverges when $\mu_I$ equals the smallest value of $\epsilon$ for which  $\langle\hat{\rho}(\epsilon)\rangle_{ 0, 0}$ is nonzero, namely $\epsilon_{\rm min}$.  On the other hand, $\chi_{\rm ps}^+$ diverges in the infrared (at $T=0$) when the lightest excitation with the quantum numbers of the $\pi^+$ has an energy equal to $\mu_I$.   Clearly, if the state at $T=0$ and $|\mu_I| < m_\pi$ were unchanged from the ordinary vacuum, this lightest excitation is simply $m_\pi$.  However, condition i) implies that in the absence of  a first order transition, the system does remain in the vacuum state until $\mu_I =\epsilon_{\rm min}$.  Together these imply that $\epsilon_{\rm min} = m_\pi$ and condition ii) is established.

In summary, the isospin Silver Blaze problem is resolved through the study of the spectrum of $\gamma_0$ times the usual Dirac operator.  As $T \rightarrow 0$, the functional determinant for any gauge configuration  is unchanged from its $\mu_I=0$ value unless $\frac{|\mu_I|}{2}$ exceeds the smallest value of $|\epsilon_j|$, where $\epsilon_j$ is the real part of the eigenvalue of the operator.  The resolution depends on the relevant gauge configurations at $T=0$ having a minimum $|\epsilon_j|$ greater than $m_\pi/2$.  It was shown that this automatically happens in QCD unless there is a first-order phase transition.  It would be interesting to see this picture emerging in lattice simulations via studies of the spectrum of $\gamma_0$ times the usual Dirac operator.

Finally, as noted above there are analogous Silver Blaze problems associated with other chemical potentials.  Of particular interest is the one associated with the baryon number: it is not plausible that we can understand nuclear matter in terms of QCD if we do not understand what happens below the phase transition to nuclear matter.  The baryon Silver Blaze problem for small chemical potentials ({\it i.e} for $\mu_B < \frac{3 m_\pi}{2}$) is easily understood via the same methods as used here.  However, the baryon Silver Blaze problem persists for all $\mu_B < M_N - B \approx 923 {\rm MeV}$ where $M_N$ is the nucleon mass and $B$ the binding energy per nucleon of nuclear matter.  Moreover, in the regime, $M_N - B > \mu_B > \frac{3 m_\pi}{2}$, the functional determinants of the relevant configurations are altered from their $\mu_I=0$ values.  Presumably, in this regime there are cancellations in the functional integrations due to the presence of the phase factors in eq.~(\ref{finaldetform}) which precisely compensate for the increases in the magnitude of the determinants.  The understanding of how this comes about represents an essential challenge to resolving the baryon Silver Blaze problem and thereby gaining insight into the first-order phase transition to nuclear matter.

\acknowledgments
The author gratefully acknowledges the support of the U.S.~Department of Energy through grant DE-FG02-93ER-40762.


\begin{thebibliography}{99}

\bibitem{ss}   D.T. Son, M.A. Stephanov,  Phys.~Rev.~Lett. {\bf 86} (2001) 592;  Phys.~Atom.~Nucl. {\bf 64} (2001) 834-; Yad.~Fiz.~{\bf 64} (2001) 899.
\bibitem{AlfKapWil99} M.~Alford, A.~Kapustin and F.~Wilczek,
Phys.~Rev.~{\bf D59}, 054502 (1999).
 \bibitem{Coh03}Thomas D.~Cohen, hep-ph/0304024, to appear in Phys.~Rev. Lett.
 \bibitem{rm}G.~ Akemann and T.~Wettig, hep-lat/0301017; G.~ Akemann, Phys.~Rev.~Lett.{\bf ~89} (2002) 072002;   J.~J.~M.~ Verbaarschot, T.~ Wettig  Ann.~Rev.~Nucl.~Part.~Sci.~ {\bf 50} (2000) 343;  P.~H.~ Damgaard, J.~C.~ Osborn, D.~ Toublan, J.~J.~M.~ Verbaarschot  Nucl.~Phys.~ {\bf B547} (1999) 305: D.~ Toublan, J.~J.~M.~ Verbaarschot, Nucl.~Phys.~ {\bf B560} (1999) 259-282. M.~A.~ Halasz, A.~D.~ Jackson, J.~J.~M.~ Verbaarschot,  Phys.~Rev.~{\bf  D56} (1997) 5140; H.~ Markum, R.~ Pullirsch, T.~ Wettig  Phys.~Rev.~Lett.~ {\bf 83} (1999) 484; G.~ Akemann, P.~ H.~ Damgaard, U.~ Magnea, S.~ Nishigaki,  Nucl.~Phys.~ {\bf B487} (1997) 721; M.~A.~ Stephanov Phys.~Rev.~Lett. {\bf 76} (1996) 4472.
\bibitem{lat}There is a vast literature on this.  The development of domain wall fermions (D.~B.~ Kaplan, Phys.~Lett.~{\bf B288} (1992) 342.) and overlap fermions ( R.~ Narayanan and H.~Neuberger,  Nucl.~Phys. ~{\bf B412} (1994) 574-606.) have stimulated numerous studies of the eigenvalues of the lattice dirac operator and its connection to chirality.

\bibitem{bc}  T.~Banks and A.~Casher, Nucl.~Phys.~ {\bf B169} (1980) 103.

\bibitem{Glasgow}  I.~M.~Barbour, C.~T.~H.~Davies and Z.~Sabeur, Phys.~Lett.~{\bf B215} (1988) 567;  I.~M.~Barbour and Z.~Sabeur, Nucl.~Phys.~{\bf B342} (19900)269;  I.~M.~Barbour and A.~J.~Bell, Nucl.~Phys.~{\bf B372} (1992) 385; A.~Hasenfratz and D.~Toussaint Nucl.~Phys.~{\bf B371} (1992) 539;  I.~M.~Barbour, S.~E.~Morrison, E.~G.~Klepfish, J.~B.~Kogut, M.-P.~Lombardo,  Phys.Rev. {\bf D56} (1997) 7063

\bibitem{rm2} M.~A.~Halasz, J.~C.~Osborn, M.~A.~Stephanov and J.~J.~M.~Verbaarschot,  Phys.~Rev.~{\bf D61} (2000) 076005.

\bibitem{Matsubura} The usual treatment is discussed in standard texts on finite temperature field theory or many-body physics.  See for example, {\it Finite-Temperature Field Theory} by J.~I.~Kapusta, (Cambridge University Press, Cambridge 1989) or {\it Quantum Many-Particle Systems} by J.~W.~Negle and H.~Orland, (Addison-Wesley, Redwood City 1988).

\end{thebibliography}
\end{document}